\documentclass[a4paper,11pt]{article}
\usepackage{pos}
\usepackage[symbol]{footmisc}
\usepackage{physics}
\usepackage{slashed}
\renewcommand{\thefootnote}{\fnsymbol{footnote}}
\usepackage{comment}

\renewcommand{\a}{\alpha}
\renewcommand{\b}{\beta}
\renewcommand{\c}{\gamma}

\newcommand{\st}{{\fontfamily{cmss}\selectfont\text{sT}}}
\newcommand{\tran}{{\fontfamily{cmss}\selectfont\text{T}}}

\title{Supergeometric Approach to Quantum Field Theory }

\author*[a]{Viola Gattus}
\author[a]{Apostolos Pilaftsis}

\affiliation[a]{Department of Physics and Astronomy, University of Manchester,\\
Manchester M13 9PL, United Kingdom}
\bigskip

\emailAdd{viola.gattus@manchester.ac.uk}
\emailAdd{apostolos.pilaftsis@manchester.ac.uk}

\abstract{We review a recently proposed SuperGeometric (SG) approach to Quantum Field Theories (QFTs) that allow for scalar-fermion field transformations in a manifestly reparameterisation covariant manner. By adopting natural choices for the field-space supermetric, we show how a non-trivial field-space Riemannian curvature can be introduced in the fermionic sector, beyond the usual one that may have its origin in the scalar part of the theory. We~present a minimal SG-QFT model with proper non-zero fermionic curvature both in two and four spacetime dimensions. Possible future directions in further developing SG-QFTs are discussed.}

\FullConference{Corfu Summer Institute 2023 "School and Workshops on Elementary Particle Physics and Gravity" (CORFU2023)\\
 23 April - 6 May , and 27 August - 1 October, 2023\\
Corfu, Greece\\}

\begin{document}
\maketitle

\section{Introduction}

Covariant differential geometric methods have played a central role in the development of Quantum Field Theory (QFT)~\cite{DeWitt:1967ub,Honerkamp:1971sh,Ecker:1972tii,Alvarez-Gaume:1981exa,Gaillard:1985uh}. In particular,
 such differential geometric methods have been employed by Vilkovisky and DeWitt~(VDW)~\cite{Vilkovisky:1984st,DeWitt:1985sg} to address the issue of gauge-fixing parameter independence in gauge and quantum gravity theories. The VDW framework was developed further by several other authors~\cite{Barvinsky:1985an,Ellicott:1987ir,Burgess:1987zi,Odintsov:1989gz}. More recently, differential geometric formalisms were utilised to resolve the so-called quantum frame problem in cosmological single-field and multi-field inflation~\cite{Kamenshchik:2014waa,Burns:2016ric,Karamitsos:2017elm}, along with the issue of uniqueness of the path-integral measure of the VDW effective action~\cite{Finn:2019aip,Falls:2018olk} beyond the Born approximation. Geometric techniques were also employed to analyse new-physics phenomena within the framework of Effective Field Theories (EFTs) beyond the Standard Model~(SM)~\cite{Alonso:2016oah,Nagai:2019tgi,Helset:2020yio,Cohen:2021ucp,Talbert:2022unj,Helset:2022tlf}, also known as SMEFT (for a review, see~\cite{Isidori:2023pyp}).

Unlike bosons, the inclusion of fermions as {\em independent} chart variables has faced a number of theoretical difficulties and limitations. Given the linearity of the fermion kinetic terms in spacetime derivatives, it is not obvious how to define a proper rank-2 tensor that could assume the role of a metric in the fermionic field space. On the other hand, in supergravity and certain string theories~\cite{Alvarez-Gaume:1981exa,Dixon:1989fj}, the geometry of the fermionic sector is related to that of bosons by supersymmetry through the so-called K\"ahler manifold. In such theories, the number of fermionic species cannot be arbitrary but directly determined by their supersymmetric bosonic counterparts. 

In these proceedings, our aim is to show that one can go beyond the standard supersymmetric framework
and treat fermions and bosons as independent field variables.
Since fermion fields are anti-commuting Grassmannian variables, their consistent description in the path-integral configuration space requires the consideration of differential {\em supergeometry} (SG) on {\em super\-manifolds}~\cite{DeWitt:2012mdz}. Recent studies~\cite{Finn:2020nvn,Pilaftsis:2022las,Gattus:2023gep} have made considerable progress in formulating reparameterisation-invariant scalar-fermion theories, where the field-space metric was defined from the action. The formalism that was put forward in~\cite{Finn:2020nvn,Pilaftsis:2022las} not only enabled one to obtain earlier known results of the effective action at the one-loop level, but also a new expression for the complete SG effective action at the two-loop order. Nevertheless, definite models with non-zero fermionic curvature have not been presented in the initial work~\cite{Finn:2020nvn}. 

In~\cite{Gattus:2023gep}, we have been able to formulate minimal SG-QFT models for the first time that feature non-zero fermionic curvature both in two and four spacetime dimensions. In formulating these minimal models, we paid attention to the issue of uniqueness in defining from the action the field-space metric of the underlying supermanifold, ${}_\alpha G_\beta$, which is also termed  {\em supermetric}. However, a proper rank-2 tensor 
can be constructed from the model-function that appears in the kinetic term of the fermions, leading to a supermetric of the field-space supermanifold which is {\em supersymmetric}. This means that~${}_\alpha G_\beta$ should be invariant under the operation of supertransposition (\text{\sf sT}) to be defined in Section 2. Furthermore, we have shown that a scalar field alone cannot induce a non-trivial field-space Riemannian curvature in
the fermionic sector. 

The remainder of the proceedings is laid out as follows. 
In Section~\ref{sec:SG}, we review the basic covariant structure of scalar-fermion SG-QFTs including their key model functions, ${}_\alpha k_\beta$ and $\zeta^\mu_\alpha$.  In the same section, we present our approach to deriving the supermetric from the classical action of an SG-QFT. In Section~\ref{sec:Models}, we show a no-go theorem for the generation of a non-zero super-Riemannian curvature in a bilinear kinetic fermionic sector from the existence of a scalar field only in SG-QFTs. In addition, we present a minimal model that realises non-zero fermionic curvature when the model function $\zeta^\mu_\alpha$ contains non-linear fermionic terms. Our main results are summarised in Section~\ref{sec:Summary}, including an outlook for physical applications that makes use of this SG-QFT framework.

\vfill\eject

\section{Supergeometry and the Scalar-Fermion Field Space}\label{sec:SG}

We briefly review some basic aspects of differential supergeometry on the scalar-fermion field space~\cite{DeWitt:2012mdz} that are relevant to the formulation of SG-QFTs. First, we note that a set of $N$ real scalar fields and $M$ Dirac fermions describe a field-space supermanifold of dimension $(N|8M)$ in four spacetime dimensions (4D). A chart of this supermanifold may be represented by the $(N+8M)$-dimensional column vector,
\begin{equation}
   \label{eq:PhiChart}
  \boldsymbol{\Phi}\ \equiv\ \left\{\Phi^{\alpha}\right\}\ =\ \left(
  \begin{array}{c}\phi^{A}\\ \psi^{X}\\
  \overline{\psi}{}^{Y\:\tran}
  \end{array}
  \right)\,,
\end{equation}
where a Greek index  like $\a=1,2,\dots, N+8M$ labels all fields. Otherwise, we use Latin letters from the beginning of the alphabet to denote individual bosonic degrees of freedom and letters from the end to denote fermionic ones. In~analogy\- to the standard theory of manifolds, general field reparameterisations of the form,
\begin{equation}
   \label{eq:PhiDiff}
    \Phi^{\alpha}\ \rightarrow\ \widetilde{\Phi}^{\alpha}\, =\, \widetilde{\Phi}^{\alpha}(\boldsymbol{\Phi})\;,
\end{equation}
become now diffeomorphisms on the supermanifold. Notice that the class of transformations in~\eqref{eq:PhiDiff} cover any {\em ultralocal} redefinitions of the scalar and fermions fields without  introducing extra spacetime derivatives of fields like $\partial_\mu \Phi^{\alpha}$. This restriction could be relaxed by adopting Finslerian-type geometries in the field space~\cite{Finn:2019aip,Craig:2023hhp,Alminawi:2023qtf}.

The Lagrangian for a general scalar-fermion theory being invariant under field-space diffeo\-morphisms, up to second order in $\partial_{\mu} \Phi^{\alpha}$, can be expressed in terms of three model functions: (i)~a rank-2 field-space tensor $_{\alpha} k_{\beta}(\boldsymbol{\Phi})$, (ii)~a mixed spacetime and field-space vector $\zeta_{\alpha}^{\mu}(\boldsymbol{\Phi})$, and (iii)~a zero-grading scalar~$U(\boldsymbol{\Phi})$ describing the potential and Yukawa interactions. Such a diffeomorphically- or frame-invariant Lagrangian reads \cite{Finn:2020nvn}
\begin{equation}
   \label{eq:LSGQFT}
    \mathcal{L}\ =\ \frac{1}{2} g^{\mu \nu} \partial_{\mu} \Phi^{\alpha}\:_{\alpha} k_{\beta}(\boldsymbol{\Phi})\: \partial_{\nu} \Phi^{\beta}\: +\: \frac{i}{2}\, \zeta_{\alpha}^{\mu}(\boldsymbol{\Phi})\: \partial_{\mu} \Phi^{\alpha}\: -\: U(\boldsymbol{\Phi}).
\end{equation}
In~\eqref{eq:LSGQFT}, $_{\alpha} k_{\beta}$
vanishes when the indices $\alpha$ or $\beta$ are fermionic, i.e.~${}_X k_A = {}_A k_Y = {}_X k_Y = 0$. Note that $_{\alpha} k_{\beta}$~plays the role of the field-space metric \cite{Finn:2019aip} for a purely bosonic theory. Therefore, the function $\zeta_{\alpha}^{\mu}$ is introduced to describe the fermionic sector. Notice that $\zeta_{\alpha}^{\mu}$ may also be used to include chiral fermions by decomposing each Dirac fermion into pairs of Majorana fermions.

The model functions, $_{\alpha} k_{\beta}$ and $\zeta_{\alpha}^{\mu}$, can unambiguously be extracted from the Lagrangian according to the following prescription~\cite{Finn:2020nvn}:
\begin{equation}
   \label{eq:kzeta}
{ }_{\alpha} k_{\beta}\ =\ \frac{g_{\mu \nu}}{D} \frac{\overrightarrow{\partial}}{\partial\left(\partial_{\mu} \Phi^{\alpha}\right)} \mathcal{L} \frac{\overleftarrow{\partial}}{\partial\left(\partial_{\nu} \Phi^{\beta}\right)}\:, \qquad 
\zeta_{\alpha}^{\mu} =\frac{2}{i}\left(\mathcal{L}-\frac{1}{2} g^{\mu \nu} \partial_{\mu} \Phi^{\gamma}{}_\gamma k_{\delta}\, \partial_{\nu} \Phi^{\delta}\right) \frac{\overleftarrow{\partial}}{\partial\left(\partial_{\mu} \Phi^{\alpha}\right)}\; .
\end{equation}
To equip the supermanifold with a metric, we need to construct a proper field-space covector~$\zeta_\alpha$ 
from~$\zeta_{\alpha}^{\mu}$. As it is evident from~\eqref{eq:LSGQFT} and explained in~\cite{Finn:2020nvn}, the Lorentz index $\mu$ in $\zeta_{\alpha}^{\mu}$ can only arise from the presence of a $\gamma^\mu$-matrix, or a $\sigma^\mu = (\sigma^0\,,\boldsymbol{\sigma})$ matrix in the chiral basis, where~$\boldsymbol{\sigma} = (\sigma^1\,, \sigma^2\,, \sigma^3)$ are the three Pauli matrices.

To find the metric of the field-space supermanifold, it proves useful to distinguish two categories of SG-QFTs depending on the actual structure of the model function~$\zeta_{\alpha}^{\mu}$. In the first category, the fermionic components of $\zeta_{\alpha}^{\mu}$ may be expressed in a factorisable form as
\begin{equation}
  \label{eq:zetagamma}
    \zeta^{\mu}_{\:\a}=\zeta_{\:\b}\: ^{\b}(\Gamma^{\mu})_{\:\a},\qquad \text{where}\quad \Gamma^{\mu}  =\left(\begin{array}{cc}
\gamma^{\mu} & 0 \\
0 & \left(\gamma^{\mu}\right)^{\tran}
\end{array}\right).
\end{equation}
The second category of SG-QFTs does not possess the factorisation property~\eqref{eq:zetagamma}. As we will see in Section~\ref{sec:Models}, the distinction between factorisable and non-factorisable $\zeta_{\alpha}^{\mu}$ affects the geometric properties of the field-space supermanifold. For the first category of SG-QFTs, it is straightforward to project a proper field-space covector $\zeta_{\:\a}$ from $\zeta_{\alpha}^{\mu}$ given in \eqref{eq:zetagamma}. The simplest way would be to introduce a differentiation with respect to the $\gamma^\mu$ matrix as done in~\cite{Finn:2020nvn}, i.e. 
\begin{equation}
   \label{eq:zeta}
    \zeta_{\alpha}\: =\: \frac{1}{D}\, \frac{\delta \zeta_{\alpha}^{\mu}}{\delta \gamma^{\mu}}\; ,
\end{equation}
where $D$ is the number of space-time dimensions. But for the second category of SG-QFTs for which $\zeta^{\mu}_{\:\a}$ does not obey~\eqref{eq:zetagamma}, one may alternatively use the more natural projection operation,
\begin{equation}
   \label{eq:projzeta}
     \zeta_{\beta}^{\mu} \:\:^\b\left(\overleftarrow{\Sigma}_{\mu}\right)_{\alpha}\: =\: \zeta_{\alpha}\;, \qquad \text{where}\quad 
  \overleftarrow{\Sigma}_{\mu}  =\frac{1}{D}\left(\begin{array}{cc}
\frac{\overleftarrow{\partial}}{\partial \gamma^{\mu}} & 0 \\
0 & \Gamma_{\mu}
\end{array}\right)\;.
\end{equation}
In the above, the differentiation acting on the fermionic components of $\zeta_{\alpha}^{\mu}$ is replaced by contraction with $\gamma^{\mu}$ matrices. In this way, the spin-$3/2$ degrees of freedom (dofs) contained in~$\zeta^{\mu}_{a}$ are projected onto spin-$1/2$ dofs in $\zeta_a$. In this study we adopt the projection method~\eqref{eq:projzeta} which can be applied to both categories of SG-QFTs.  

An important geometric property of SG-QFTs as described by the Lagrangian~$\mathcal{L}$ in \eqref{eq:LSGQFT} is that $\mathcal{L}$ is a scalar in the field-space supermanifold.  In other words, $\mathcal{L}$ remains invariant under the field redefinitions in~\eqref{eq:PhiDiff}, provided all model functions and the field-space tangent vectors are appropriately transformed. In this SG framework, we have 
\begin{equation}
    \partial_{\mu}\widetilde{\Phi}^\a(\boldsymbol{\Phi})\: =\: \partial_{\mu}\Phi^\b(\boldsymbol{\Phi}) \:_{\b }J^\a(\boldsymbol{\Phi})\;,
\end{equation}
where $\:_{\b }J^\a =\:_{\b,}\widetilde{\Phi}^{\a}$ is the Jacobian of the transformation and the subscript $\beta$ before the comma on the left side of $\widetilde{\Phi}^\a$ denotes ordinary left-to-right differentiation with respect to the field $\Phi^\b$.

A field-space supermanifold of interest to us must be endowed with a rank-2 field-space tensor~$_\a G_\b$, which is supersymmetric, i.e. 
\begin{equation}
    {}_\alpha G_\beta\: =\: ({}_\alpha G_\beta)^{\sf sT}\: =\: (-1)^{\alpha +\beta +\alpha\beta} {}_\beta G_\alpha\;,
\end{equation}
and non-singular. Such a supermanifold is called Riemannian~\cite{DeWitt:2012mdz} and the rank-2 field-space tensor~$_\a G_\b$ is known as the \textit{supermetric}. Its inverse $^\a G^\b$, deduced from the identity: $ _\a G_\c \:G^{\c \b}=\: _\a\delta^{\b}$, satisfies
\begin{equation}
    ^\a G^\b\: =\: G^{\a \b}\: =\: (-1)^{\a \b}\: G^{\b \a} .
\end{equation}
In the above, we have employed the compact index calculus and conventions by DeWitt in~\cite{DeWitt:2012mdz}, so that the exponents of $(-1)$ determine the grading of the respective quantities and take the values~$0$ or~$1$ for commuting or anticommuting fields, respectively. According to  DeWitt's conventions, the usual tensor contraction between indices can only be performed if the two indices to be summed over are adjacent. Otherwise, extra factors of $(-1)$ must be introduced whenever two indices are swapped.

Given the supermetric ${}_\alpha G_\beta$, the Christoffel symbols
$\Gamma^{\a}_{\:\:\b \c}$ can be evaluated as in~\cite{DeWitt:2012mdz}, from which the super-Riemann tensor is obtained
\begin{equation}
   \label{eq:Riemann}
R^\a_{\:\:\b \c \delta}\: =\: -\,\Gamma^{\a}_{\:\:\b \c,\delta}\: +\: (-1)^{\c \delta}\: \Gamma^{\a}_{\:\:\b \delta,\c}\: +\:(-1)^{\c(\sigma+\b)}\: \Gamma_{\:\:\sigma \c}^{\a} \Gamma_{\:\:\b \delta}^{\sigma}\: -\: (-1)^{\delta(\sigma+\b+\c)}\: \Gamma_{\:\:\sigma \delta}^{\a} \Gamma_{\:\:\b \c}^{\sigma}\; .
\end{equation}
The super-Ricci tensor is obtained by contracting the first and third indices of the super-Riemann tensor~\cite{DeWitt:2012mdz},
\begin{equation}
    R_{\a \b}\: =\: (-1)^{\c (\a + 1)}R^{\c}_{\:\:\a \c \b}\; .
\end{equation}
 Further contraction of the remaining two indices of $R_{\a \b}$ yields the super-Ricci scalar,
\begin{equation}
  \label{eq:Rscalar}
    R\: =\:  R_{\a \b}\:G^{\b \a}\; .
\end{equation}
Note that the super-Ricci tensor is supersymmetric, i.e.~$R_{\a \b} = (R_{\a \b})^{\sf sT} =
(-1)^{\a \b}R_{\b \a}$.

To determine the supermetric $_\a G_\b$ of the scalar-fermion field space, we follow the procedure presented in~\cite{Finn:2020nvn, Pilaftsis:2022las}. After calculating the projected model function~$\zeta_\a$ as stated in~\eqref{eq:projzeta}, we may now construct the rank-2 field-space anti-supersymmetric tensor
\begin{equation}
  \label{eq:lambda}
   {}_{\alpha} \lambda_{\beta}\ =\ \frac{1}{2}\, \Big({}_{\a,}\zeta_{\beta}\: -\: (-1)^{\alpha+\beta+\alpha \beta}  \:_{\b,}\zeta_{\alpha}\Big)\; .
\end{equation}
Exactly as happens for the anti-symmetric field strength tensor $F_{\mu \nu}$ in QED in curved spacetime, the derivatives appearing in \eqref{eq:lambda} are ordinary derivatives and not covariant ones, since the Christoffel symbols drop out for such constructions of anti-supersymmetric rank-2 tensors like~$_\a \lambda_\b$.

The so-constructed ${ }_{\alpha} \lambda_{\beta}$ turns out to be singular in the presence of scalar fields, and so the scalar contribution $_{\alpha} k_{\beta}$  has to be added which results in the new rank-2 field-space tensor,
\begin{equation}
   \label{eq:Lambda}
_{\alpha} \Lambda_{\beta}\ =\:  _{\alpha} k_{\beta} \:+\:  _{\alpha} \lambda_{\beta}\; .
\end{equation}
However, $_{\alpha} \Lambda_{\beta}$ cannot act as a supermetric, since it
is not supersymmetric. To find a suitable rank-2 tensor that satisfies the latter property, we make use of the vielbein formalism~\cite{Schwinger:1963re, Yepez:2011bw} which allows to
compute the field-space vielbeins $_\a e^a$, if the form of $_{\alpha} \Lambda_{\beta}$ is known in the local field-space frame. For the latter, we demand that the Lagrangian~\eqref{eq:LSGQFT} assumes the canonical Euclidean form in this local frame. In this way, we may compute the field-space supermetric as~\cite{Finn:2020nvn}
\begin{equation}
  \label{eq:aGb}
    _\a G _\b \:=\: _\a e^a\:\:_a H_b\:\:^be^{\st}_\b\;,
\end{equation}
where 
\begin{equation}    
   \label{eq:aHb}
{ }_{a} H_{b} \equiv\left(\begin{array}{ccc}
{\bf 1}_{N} & 0 & 0 \\
0 & 0 & {\bf 1}_{4M}  \\
0 & -{\bf 1}_{4M} & 0 
\end{array}\right)
\end{equation}
is the local field-space metric in 4D.

Finally, following the VDW formalism~\cite{Vilkovisky:1984st,DeWitt:1985sg}, we promote the field space to a configuration space, so as to take into account the spacetime dependence of the fields. In this configuration space, the coordinate charts are extended as
\begin{equation}
  \label{eq:PhiCS}
    \Phi^{\hat{\alpha}}\: \equiv\: \Phi^{\alpha}\left(x_{\alpha}\right),
\end{equation}
where $x_\a$ is the spacetime coordinate of a generic field $\Phi^\a$. Likewise,
the supermetric gets generalised as 
\begin{equation}
   \label{eq:CSGmetric}
    { }_{\hat{\alpha}} G_{\hat{\beta}}\ =\ { }_{\alpha} G_{\beta}\: \delta(x_{\alpha} - x_{\beta})\;,
\end{equation}
where $\delta(x_{\alpha} - x_{\beta})$ is the $D$-dimensional $\delta$-function.
This generalisation affects the Christoffel symbols and the Riemann tensors, as given in more detail in~\cite{Finn:2019aip}.

\section{Minimal SG-QFT Models}\label{sec:Models}

In this section we first give further clarifying details of the proof of a no-go theorem that was presented earlier in~\cite{Gattus:2023gep}. Specifically, we explain why no non-zero field-space curvature can be generated in the fermionic sector from a single scalar field and multiple fermions, as long as the model function $\zeta^\mu_\alpha$ only contains linear terms in the fermionic fields. We then present a minimal SG-QFT model with non-zero fermionic curvature which is induced by including non-linear terms in fermion fields in $\zeta^\mu_\alpha$.

\subsection{No-Go Theorem for Fermionic Field-Space Curvature}

The simplest scenario with a single boson $\phi$ and one Dirac fermion $\psi$ was considered in~\cite{Finn:2020nvn}, where it was shown that this case reduces to a flat field space.
Here, we will analyse a more general scenario with a single boson $\phi$ and a multiplet
$\boldsymbol{\psi} =\{ \psi^X\} $ of Dirac fermion fields  (with $X=1,2,\dots M$). In 4D, such a scenario has $(1|8M)$ field-space coordinates. Up to second order in spacetime derivatives, the Lagrangian for such a system is given  by
\begin{equation}
   \label{eq:NoGo}
\begin{aligned}
\mathcal{L}\ =&\ \frac{1}{2}\,k(\phi)\, (\partial^{\mu} \phi)\, (\partial_{\mu} \phi) \: -\: \frac{1}{2} h_{X Y}(\phi)\, \overline{\psi}^{X} \gamma^{\mu} \psi^{Y} (\partial_{\mu} \phi) \\
&+\: \frac{i}{2}\, g_{X Y}(\phi)\left[\overline{\psi}^{X} \gamma^{\mu} (\partial_{\mu} \psi^{Y})\, -\, (\partial_{\mu} \overline{\psi}^{X}) \gamma^{\mu} \psi^{Y}\right]\;. 
\end{aligned}
\end{equation}
Evidently, the single field $\phi$ cannot induce by itself a non-zero Riemannian curvature in the scalar sector. Consequently, if a non-trivial field-space curvature exists, this can only come from the fermionic sector of the Lagrangian~\eqref{eq:NoGo}.

In the following, we will show that the addition of multiple fermions as described by the Lagrangian~${\cal L}$ in~\eqref{eq:NoGo} will still give rise to a flat field space. 
But before doing so, we will examine the conditions under which a field transformation can bring the model function $\zeta_\a$ as well as ${\cal L}$ into a canonical Cartesian form.
To this end, let us consider the following redefinition of fermionic fields:
\begin{equation}
   \label{eq:psitilde}
    \boldsymbol{\psi}\ \longrightarrow\ \widetilde{ \boldsymbol{\psi}}\, =\, \boldsymbol{K}(\phi)^{-1}\,\boldsymbol{\psi}\;,
\end{equation}
where $\boldsymbol{K}$ is a $4M\times 4M$-dimensional matrix that 
only depends on the scalar field~$\phi$. The field reparameterisation~\eqref{eq:psitilde} modifies the fermionic part of the Lagrangian \eqref{eq:NoGo} according to the transformations,
\begin{equation}
   \label{eq:Jacobian_NoGo}
    \partial_\mu \widetilde{\Phi}^\a\ =\ \partial_\mu \Phi^\b\: _\b J^\a\;=\:^\a J^\st_\b\: \partial_\mu\Phi^\b\:,\qquad 
    \widetilde{\zeta}_\a\ =\ \zeta_\b\: ^\b (J^{-1})^{\st}_{\a}\;,
\end{equation}
where $\{\zeta_\a\}=\big(i\boldsymbol{\overline{\psi}}\,\boldsymbol{h}\,\boldsymbol{\psi}\:,\:\boldsymbol{\overline{\psi}}\,\boldsymbol{g}\:,\:\boldsymbol{\psi}{}^\tran\boldsymbol{g}{}^\tran\big)$.
For the specific reparametrisation in \eqref{eq:psitilde}, the Jacobian matrix $\boldsymbol{J}^\st =\{{}^\a J^\st_\b\}$ and its inverse, read
\begin{equation}
  \boldsymbol{J}^\st=  \left(\begin{array}{ccc}
        1 & 0 & 0\\
        (\boldsymbol{K}^{-1})^\prime\:\boldsymbol{\psi} & \boldsymbol{K}^{-1}&0\\
        (\boldsymbol{K}^{*\:-1})^\prime\:\boldsymbol{\overline{\psi}}\:^\tran& 0&\boldsymbol{K}^{*\:-1}
    \end{array}\right)\,,\qquad 
    (\boldsymbol{J}^{-1})^\st=  \left(\begin{array}{ccc}
        1 & 0 & 0\\
        \boldsymbol{K}^\prime\:\widetilde{\boldsymbol{\psi}} & \boldsymbol{K}&0\\
        (\boldsymbol{K}^*)^\prime\:\widetilde{\boldsymbol{\overline{\psi}}}{}^{\,\tran}& 0&\boldsymbol{K}^{*}
    \end{array}\right).
\end{equation}
Here and in the following, a prime (${}^\prime$) will stand for differentiation with respect to the field $\phi$, e.g.~${\boldsymbol{K}^\prime \equiv \partial\boldsymbol{K}(\phi)/\partial\phi}$.
Imposing the requirement that the transformed model function $\widetilde{\zeta}_\a$ has the standard Cartesian form, i.e.~$\{\widetilde{\zeta}_\a\} = \big(0\:,\:\widetilde{\boldsymbol{\overline{\psi}}}\:,\:\widetilde{\boldsymbol{\psi}}\big)$,
leads to the following two conditions:
\begin{eqnarray}
   \label{eq:Cond1}
    i\,\boldsymbol{K}^{\dagger}\,\boldsymbol{h}\, \boldsymbol{K}\: +\: \boldsymbol{K}^{\dagger}\boldsymbol{g}\,\boldsymbol{K}^\prime\: -\: 
    \boldsymbol{K}^{\prime\dagger}\boldsymbol{g}\,\boldsymbol{K} &=& {\bf 0}\; ,\\
  \label{eq:Cond2}
\boldsymbol{K}^{\dagger} \boldsymbol{g}\, \boldsymbol{K} &=& {\bf 1}\;.
\end{eqnarray}
with $\boldsymbol{g} = \{g_{XY}\}$ and $\boldsymbol{h} =\{h_{XY}\}$ being Hermitian matrices.
Even though it is straightforward to find a solution to the system~\eqref{eq:Cond1} and~\eqref{eq:Cond2} for the case of a single fermion field~\cite{Finn:2020nvn}, it becomes non-trivial in the presence of many fermions. To this end, we first rescale $\boldsymbol{K}$ as follows:
\begin{equation}
  \label{eq:Ktilde}
\boldsymbol{K}\ =\  \boldsymbol{g}^{-1/2}\, \boldsymbol{V}\,,  
\end{equation}
with $\boldsymbol{g}^{1/2}\boldsymbol{g}^{1/2} = \boldsymbol{g}$ such that $[\boldsymbol{g}^{1/2},\, \boldsymbol{g}] = {\bf 0}$.
As a consequence of the above rescaling,  $\boldsymbol{V}$ should be a unitary matrix, because the second condition~\eqref{eq:Cond2} implies $\boldsymbol{V}^\dagger \boldsymbol{V} = {\bf 1}$. 

Now, after some familiarisation with the matrix expression on the LHS of~\eqref{eq:Cond1}, we require that the unitary matrix $\boldsymbol{V}$ satisfies the differential equation,
\begin{equation}
    \frac{\partial \boldsymbol{V}}{\partial\phi}\ =\ -\frac{i}{2}\boldsymbol{g}^{-1/2}\boldsymbol{h}\, \boldsymbol{g}^{-1/2}\, \boldsymbol{V},
\end{equation}
whose formal solution may be represented as
\begin{equation}
    \boldsymbol{V}(\phi)\ =\ \text{T}_{\phi}\bigg\{ \exp\left( -\frac{i}{2}\int_{0}^{\phi}\boldsymbol{g}^{-1/2}\boldsymbol{h}\, \boldsymbol{g}^{-1/2}\: d \phi\right)\;\bigg\}\,.
\end{equation}
Note that $\text{T}_{\phi}$ is the field-space equivalent of the usual time ordering operator. After some algebra, we notice that the first condition~\eqref{eq:Cond1} gets fulfilled if $\boldsymbol{g}$ and $\boldsymbol{g}^{\prime}$, commute, i.e.~$[\boldsymbol{g}\,, \boldsymbol{g}^{\prime}] = {\bf 0}$. This can only happen naturally, if either $\boldsymbol{g}$ is a constant matrix, or a flavour basis exists
for which $\boldsymbol{g}$ is diagonal for all field values~$\phi$ in the flavour space.

In this latter case, after the field transformation~\eqref{eq:psitilde}, the Lagrangian in terms of the 
transformed fields takes on the Cartesian form:
\begin{equation}
  \label{eq:LCartesian}
{\cal L}\ = \  \frac{1}{2}\,(\partial^{\mu} \phi)\, (\partial_{\mu} \phi)\: +\: \frac{i}{2}\, \left[\widetilde{\overline{\psi}}{}^{X} \gamma^{\mu} (\partial_{\mu} \widetilde{\psi}^{X})\, -\, (\partial_{\mu} \widetilde{\overline{\psi}}{}^{X}) \gamma^{\mu} \widetilde{\psi}^{X}\right]\;,    
\end{equation}
where $k(\phi) =1$ was taken here for simplicity. Here, we must remark that if we wish to simply eliminate the model-function $\boldsymbol{h}$, but keep $\boldsymbol{g}$ non-zero in general, then a different non-unitary form of $\boldsymbol{K}(\phi)$ should be utilised: $\boldsymbol{K}(\phi) = \text{T}_{\phi}\Big\{ \exp\Big(- \frac{i}{2}\int_{0}^{\phi} \boldsymbol{g}^{-1} \boldsymbol{h}\, d \phi\Big)\,\Big\}$. As demonstrated explicitly in~\cite{Gattus:2023gep}, this form of $\boldsymbol{K}(\phi)$ obeys only the first condition~\eqref{eq:Cond1}, as it should be.

It is now important to analyse the geometry of the field-space supermanifold for this theory, within our formalism of an SG-QFT. 
In particular, we use the general formula~\eqref{eq:aGb} to analytically compute the supermetric ${ }_{\alpha} G_{\beta}$ derived from $\zeta_\a$ associated with the Lagrangian ${\cal L}$ in~\eqref{eq:NoGo}. In this way, we~obtain~\cite{Finn:2020nvn, Pilaftsis:2022las}
\begin{equation}
   \label{eq:metric}
    { }_{\alpha} G_{\beta}\ =\ \left(\begin{array}{ccc}
k-\frac{1}{2}\boldsymbol{\overline{\psi}}\left(\boldsymbol{g}^{\prime}- i \boldsymbol{h}\right)\boldsymbol{g}^{-1}\left(\boldsymbol{g}^{\prime} + i \boldsymbol{h}\right)\boldsymbol{\psi} & -\frac{1}{2}\boldsymbol{\overline{\psi}}\left(\boldsymbol{g}^{\prime}- i \boldsymbol{h}\right) & \frac{1}{2}\boldsymbol{\psi}^{\:\tran}\left(\boldsymbol{g}^{\prime\: \tran}  + i \boldsymbol{h}^\tran\right)\\
\frac{1}{2}\left(\boldsymbol{g}^{\prime\:\tran}- i \boldsymbol{h}^\tran\right) \boldsymbol{\overline{\psi}}^{\:\tran} & 0 & \boldsymbol{g}^\tran 1_{4} \\
-\frac{1}{2}\left(\boldsymbol{g}^{\prime}+ i \boldsymbol{h}\right) \boldsymbol{\psi} & -\boldsymbol{g} 1_{4} & 0
\end{array}\right)\;.
\end{equation}
However, as opposed to what was conjectured in~\cite{Finn:2020nvn}, we find that the super-Riemann tensor computed from the field-space supermetric~\eqref{eq:metric} vanishes identically, thus implying that the field-space supermanifold is flat~\cite{Gattus:2023gep}\footnote{The authors of~\cite{Assi:2023zid} posit a different supermetric. Specifically, the upper left scalar entry ${}_\phi G_\phi$ has a different form from the one given in~\eqref{eq:metric}, which leads to a non-zero Riemanian tensor.}. 

If we now add more than one scalar field $\phi^A$ to the theory,
a non-zero scalar curvature can be generated that will originate as usual from the scalar-dependent model functions ${}_A k_B$ and $h_{AXY}$, with $A,B > 1$. But, in order to get a non-zero Riemann tensor from fermions only, we have to introduce non-linear powers of fermionic fields in the model function $\zeta^\mu_\alpha$. This is the essence of the No-Go theorem for proper non-zero fermionic Riemannian curvature presented in~\cite{Gattus:2023gep}. In the next subsection, we present two SG-QFT models that realise non-zero fermionic curvature.

\subsection{Minimal Model with Non-zero Fermionic Field-Space Curvature}

Having gained some insight from the discussion above on the no-go theorem, we now consider a 2D SG-QFT model, which includes one scalar field~$\phi$ and one Dirac fermion~$\psi = (\psi_1\,, \psi_2)^\tran$, and contains non-linear fermionic kinetic terms in $\psi$ or $\overline{\psi}$. More explicitly, the Lagrangian of this simple model~I reads:
\begin{equation}
   \label{eq:ModelI}
\begin{aligned}
\mathcal{L}_{\mathrm{I}}\ =&\ \frac{1}{2} k\, (\partial_{\mu}\phi)\, (\partial^{\mu} \phi)\: +\: \frac{i}{2} \left(g_0 +g_1\overline{\psi}\psi \right)\left[\overline{\psi} \gamma^{\mu} (\partial_{\mu} \psi)\,-\, (\partial_{\mu} \overline{\psi}) \gamma^{\mu} \psi\right]\: +\:
Y\,\overline{\psi} \psi\: -\: V\;,
\end{aligned}
\end{equation}
where $\gamma^\mu = (\sigma^1\,, -i\sigma^2)$. Here, all the model functions~$k$, $g_0$,  $g_1$, $Y$ and $V$ depend on the scalar field~$\phi$. Note that the model function $\zeta^\mu_\alpha$ derived from~\eqref{eq:ModelI} takes on the factorisable form of~\eqref{eq:zetagamma}, with
\begin{equation}
   \label{eq:zetaMI}
\zeta_\a\ =\ \Big\{0\,, \big(g_0 + g_1\overline{\psi}\psi\big)\,\overline{\psi}\,, \big(g_0 + g_1\overline{\psi}\psi\big)\,\psi^{\tran}\Big\}\;.
\end{equation}
Using the method of~\cite{Finn:2020nvn} briefly outlined in Section~2, we may derive the field-space super\-metric~$\boldsymbol{G} = \{ {}_\alpha G_\beta\}$ in the superspace $\boldsymbol{\Phi}^\tran = (\phi\,, \psi^{\tran}\,, \overline{\psi})$,
\begin{equation}
   \label{metric_2d_general}
    \boldsymbol{G}\, =\,\left(\begin{array}{crr}
k\: +\: b^\tran (d^{-1})^\tran a^\tran -\, a\,d^{-1}\,b\: 
& \ -a &\ b^\tran\\
\hspace{3mm} a^\tran &\ 0 &\ d^\tran\\
-b &\ -d & 0\hspace{2mm}{}
\end{array}\right),
\end{equation}
where
\begin{equation}
\begin{aligned}
a\ =\ \frac{1}{2}\,\overline{\psi}\left(g_0^\prime + g_1^\prime \overline{\psi}\psi\right)\;,\quad
b\ =\ \frac{1}{2}\left(g_0^\prime + g_1^\prime \overline{\psi}\psi\right)\psi\; ,\quad
d\ =\ \left(g_0+ g_1\overline{\psi}\psi\right) 1_2\, +\, g_1\psi\overline{\psi}\; ,
\end{aligned}
\end{equation}
and a prime $({}^\prime)$ on the model functions~$g_{0,1}$ denotes differentiation with respect to $\phi$. Note that $\boldsymbol{G}$ is supersymmetric, since $b^\tran (d^{-1})^\tran a^\tran = -a\,d^{-1} b$. 

Given the supermetric~$\boldsymbol{G}$, we may now compute the  
non-zero components of the Riemann tensor. For instance, if
$g_0=g_1=1$, these components are found to be
\begin{equation}
   \label{eq:RiemannMI}
\begin{aligned}
R^{\psi_1}_{\:\:\psi_1\overline{\psi}_1\psi_2}&=-R^{\overline{\psi}_2}_{\:\:\overline{\psi}_2\overline{\psi}_1\psi_2}= \psi_1\overline{\psi}_2-1\;, \\
 R^{\psi_1}_{\:\:\psi_1\overline{\psi}_2\psi_2}&=R^{\overline{\psi}_1}_{\:\:\overline{\psi}_2\overline{\psi}_1\psi_2}=-R^{\overline{\psi}_1}_{\:\:\overline{\psi}_1\overline{\psi}_2\psi_2}= -\psi_1\overline{\psi}_1\;, \\
 R^{\psi_2}_{\:\:\psi_1\overline{\psi}_1\psi_2}&= R^{\overline{\psi}_2}_{\:\:\overline{\psi}_2\overline{\psi}_1\psi_1}=-R^{\psi_2}_{\:\:\psi_2\overline{\psi}_1\psi_1} =\psi_2\overline{\psi}_2\;, \\
 R^{\psi_2}_{\:\:\psi_1\overline{\psi}_2\psi_2}&=R^{\overline{\psi}_1}_{\:\:\overline{\psi}_1\overline{\psi}_2\psi_1}=-R^{\psi_2}_{\:\:\psi_2\overline{\psi}_2\psi_1}=-R^{\overline{\psi}_1}_{\:\:\overline{\psi}_2\overline{\psi}_1\psi_2}= 1-\psi_2\overline{\psi}_1\;.
\end{aligned}
\end{equation}
Hence, the minimal SG-QFT model of~\eqref{eq:ModelI} exhibits a non-zero fermionic field-space curvature. Allowing for $\phi$-dependent model functions $g_{0,1}$, the super-Ricci scalar evaluates to
\begin{equation}
  \label{eq:RicciMI}
R\ =\ \frac{4g_1}{g_0^2}\: +\: \left(\frac{2 g_1 g_0^{\prime} g_1^{\prime}}{g_0^3 k}-\frac{2
   g_1^2 g_0^{\prime\:2}}{g_0^4 k}-\frac{g_1^{\prime\:2}}{2 g_0^2 k}\right)(\overline{\psi}\psi)^2\;.
\end{equation}
Observe that $R$ is a Lorentz scalar, but not a real-valued expression due to the appearance of the fermionic bilinear term $(\overline{\psi}\psi)^2$. For $g_0=g_1=1$, the super-Ricci scalar simplifies to
\begin{equation}\label{eq:RicciMI0}
 R\ =\ 4\;.
\end{equation}
It is important to remark here that the same result~\eqref{eq:RicciMI0} would have been obtained in the absence of the bosonic field~$\phi$. Consequently, the non-vanishing field-space curvature arises from the non-linear terms in the fermion fields in $\zeta_\alpha$ through the model function $g_1$ in~\eqref{eq:zetaMI}.

The above consideration can be easily extended to a 4D version of the SG-QFT Model~I considered in~\eqref{eq:ModelI}. In this case,
$\gamma^\mu$ stand for the usual 4D Dirac matrices, and the 
Dirac fermion has four components: $\psi^\tran = (\psi_1\,, \psi_2\,, \psi_3\,, \psi_4)$. The 4D SG-QFT model has $(1|8)$ dimensions giving rise to rather lengthy expressions for the super-Riemann tensor, which we will not present here. Instead, we give the field-space super-Ricci scalar, 
\begin{equation}
   \label{eq:RicciMI4D}
\begin{aligned}
R\ =&\ \frac{24 g_1}{g_0^2}\,-\, \frac{24g_1^2}{g_0^3}(\overline{\psi}\psi) \,+\,\left(\frac{2 g_1 g_0^\prime g_1^\prime}{g_0^3 k}-\frac{2 g_1^2 g_0^{\prime\:2}}{g_0^4 k}-\frac{g_1^{\prime\:2}}{2 g_0^2 k}-\frac{4 g_1^3}{g_0^4}\right) (\overline{\psi}\psi)^2\\
 &\ +\left(-\frac{16 g_1^2 g_0^\prime g_1^\prime}{g_0^4 k}+\frac{16 g_1^3 g_0^{\prime\:2}}{g_0^5 k}+\frac{4
   g_1 g_1^{\prime\:2}}{g_0^3 k}+\frac{40 g_1^4}{g_0^5}\right) (\overline{\psi}\psi)^3\\
&\ +\left(\frac{80 g_1^3 g_0^\prime g_1^\prime}{g_0^5 k}-\frac{80 g_1^4 g_0^{\prime\:2}}{g_0^6 k}-\frac{20
   g_1^2 g_1^{\prime\:2}}{g_0^4 k}+\frac{20 g_1^5}{g_0^6}\right) (\overline{\psi}\psi)^4\; .
\end{aligned}
\end{equation}
For $g_0=g_1=1$, the field-space Ricci scalar takes on the simpler form,
\begin{equation}
   \label{eq:RicciMI04D}
    R\ =\ 24\: -\: 24\,(\overline{\psi}\psi)\: -\: 4\,(\overline{\psi}\psi)^2\: +\: 40\,(\overline{\psi}\psi)^3\: +\: 20\,(\overline{\psi}\psi)^4\; .
\end{equation}
We note that \eqref{eq:RicciMI04D} becomes identical to the result one would obtain in a system with two fermions in 2D. This should be expected, since the number of degrees of freedom and the structure of the Lagrangian~\eqref{eq:ModelI} are exactly the same for the two cases. 

Let us now discuss an important feature of the geometric construction of Lagrangian~\eqref{eq:ModelI}, and SG-QFTs in general. Specifically, one  may  notice that under a naive non-linear reparameterisation of the fermion fields,
\begin{equation}
  \label{eq:Nlinfermion}
    \widetilde{\psi}\: =\: \psi \:\sqrt{1+\overline{\psi}\psi}\;,\qquad  \widetilde{\overline{\psi}}\: =\: \sqrt{1+\overline{\psi}\psi}
\:\overline{\psi} \;,
\end{equation}
one can turn a standard (canonical) Dirac Lagrangian,
\begin{equation}
   \label{eq:LDirac}
    \mathcal{L}_{\text{D}}\ =\ \frac{i}{2}\,\left[\widetilde{\overline{\psi}}\,(\slashed{\partial}\widetilde{\psi})\: -\:(\slashed{\partial}\widetilde{\overline{\psi}})\,\widetilde{\psi}\right]\,,
\end{equation}
into the Lagrangian~\eqref{eq:ModelI}, in which $g_0 = g_1 = 1$ and all remaining model functions are set to zero, $k = Y = V = 0$.
This would seem to suggest that  a curved field-space theory can be obtained from a flat one by means of a non-linear reparameterisation like~\eqref{eq:Nlinfermion}, and vice-versa. 

However, within our SG-QFT framework, such a transformation by itself does not dictate the geometry of the field-space supermanifold. As explained in the previous section, the supermetric $\boldsymbol{G}$ of the field-space is determined by the kinetic model function $\zeta_\a$ through the steps leading to formula~\eqref{eq:aGb}. It was therefore crucial to show that the so-derived supermetric $\boldsymbol{G}$ given in~\eqref{metric_2d_general} leads to a non-zero Riemannian tensor 
implying a non-zero fermionic curvature. Consequently, it is important to emphasise here that  different non-standard non-linear forms of the model function $\zeta_{\alpha}^{\mu}$ are expected to provide distinct supergeometric constructions of Lagrangians, involving different supermetrics $\boldsymbol{G}$ with non-zero super-Riemanian curvature. The superdeterminant of $\boldsymbol{G}$ determines the path-integral measure which in turn impacts the effective action beyond the classical approximation~\cite{Finn:2019aip,Finn:2020nvn}.

In addition to the minimal model discussed above, we may formulate SG-QFT models that can have a richer geometric structure, such as the one~based on the Lagrangian~\cite{Gattus:2023gep}:
\begin{equation}
   \label{eq:MII}
    \mathcal{L}_{\rm II}\ =\ \frac{i}{2}\left[\overline{\psi}\gamma^{\mu}(\partial_{\mu}\psi)-(\partial_{\mu}\overline{\psi})\gamma^{\mu}\psi\right]\: +\: \frac{i}{2}\overline{\psi}\gamma^{\mu}\psi\left[\overline{\psi}(\partial_{\mu}\psi)-(\partial_{\mu}\overline{\psi})\psi\right]\;,
\end{equation}
with $\zeta^{\mu}_\a\:=\big( 0\:,\:\overline{\psi}\gamma^{\mu}+(\overline{\psi}\gamma^{\mu}\psi)\,\overline{\psi}\:,\:\psi^\tran\gamma^{\mu\:\tran}+(\overline{\psi}\gamma^{\mu}\psi)\,\psi^\tran\big)$.
Note that naive redefinitions of the fermion fields: $\widetilde{\psi} = \psi\, f(\overline{\psi}\psi )$ and $\widetilde{\overline{\psi}} =  f^*(\overline{\psi}\psi )\,\overline{\psi}$, where $f$ is some judicious function like~\eqref{eq:Nlinfermion}, cannot bring the Lagrangian~\eqref{eq:MII} into a canonical form like~\eqref{eq:LDirac}.
The interested reader may find more details of Model~II in~\cite{Gattus:2023gep}.

We end this section by commenting on the flavour covariance of an SG-QFT with many species of fermions. An equivalent class of Lagrangians can be consistently constructed through flavour field redefinitions, $\boldsymbol{\psi} \to \boldsymbol{\widetilde{\psi}} = \boldsymbol{U} \boldsymbol{\psi}$,
where  $\boldsymbol{U}$ is a unitary flavour-rotation matrix that does not depend on the scalar fields. The new supermetric in the flavour-transformed basis is derived from the usual rank-2 covariance relation,
$_\a \widetilde{G}_\b =\:_\a (J^{-1})^\c\:_\c G_\delta \:^\delta (J^{-1})^\st_\b\; .$
The so-derived supermetric can be shown to be equivalent to the supermetric that would be obtained by extracting the new model functions from a flavour-transformed Lagrangian${}$, e.g.~$\mathcal{\widetilde{L}}$. 
This last property provides further support of the mathematical consistency of our SG-QFT framework.

\section{Summary and Outlook}\label{sec:Summary}

We critically reviewed the frame-covariant formalism presented in~\cite{Finn:2020nvn} and~\cite{Gattus:2023gep} on scalar-fermion theories. We illustrated how the scalar and fermion fields define a coordinate system (also called chart) that describes a supermanifold in the configuration space of the respective QFT.  We discussed the issue of uniqueness of the supermetric, explaining how different choices of the latter lead to distinct Supergeometric QFTs in the off-shell kinematic region, and beyond the tree level. 

Given a self-consistent choice for the supermetric, we have demonstrated that scalar fields~$\phi^A$ alone do not provide a new source of curvature in the fermionic sector beyond the one that originates from the scalar-dependent model functions ${}_A k_B$, and $h_{AXY}$, once more than one scalar field with $A,B > 1$ are added to the theory. In particular, we have shown that the fermionic curvature vanishes for a scalar-dependent $g_{XY}(\phi)$, even if $h_{XY}(\phi)$ (with $A=1=\phi$) is non-zero. To avoid this No-Go theorem, we have to introduce non-linear powers of fermionic fields in the model function $\zeta^\mu_\alpha$ which can give rise to non-zero fermionic curvature, implying a non-zero super-Riemann tensor. In this way, we were able to formulate a minimal SG-QFT model that realises non-zero fermionic curvature both in two and four spacetime dimensions up to second order in spacetime derivatives. We should stress here that the resulting super-Riemann tensor and super-Ricci scalar may contain fermionic bilinears which are no proper real numbers. This should be contrasted with Supergravity theories~\cite{Alvarez-Gaume:1981exa} where the curvature is a real-valued expression dictated by the scalar part of the Kaehler manifold, on which the fermions were treated as tangent vectors. 

\vfill\eject

Several new research directions open up within the framework of SG-QFTs. Most notably, we expect that SG-QFTs will lead to a complete geometrisation of realistic theories of micro-cosmos, such as the SM and its gravitational sector. We also envisage that SG-QFTs will provide a new portal to the dark sector, where dark-sector fermionic fields will change the dispersion properties of weakly interacting particles, like SM neutrinos and axions. Our plan is to investigate some of the above issues in future works.

\subsection*{Acknowledgements} 

The work of AP is supported in part by the STFC Research Grant ST/X00077X/1. VG~acknow\-ledges support by the University of Manchester through the President's Doctoral Scholar Award. 

\vfill\eject

\end{document}